# Two-Person Cooperative Games with δ-Rationality

Fang-Fang Tang, Peking University (fftang2012@126.com)
& Yongsheng Xu Georgia State University (yxu3@gsu.edu)

June 19, 2025

**Abstract.** A player's payoff is modeled as consisting of two parts: a rational-value part and a distortion-value part. It is argued that the (total) payoff function should be used to explain and predict the behaviors of the players, while the rational value function should be used to conduct welfare analysis of the final outcome. We use the Nash demand game to illustrate our model.

**Acknowledgment.** We thank Mr. Dahong Liu (a techno-entrepreneur) and Mr. Rickey Di Cao (a fund manager) for sharing their personal business experiences including awfully unpleasant real cases with sad consequences. We would like to devote this work to late Professors John C. Harsanyi and Reinhard Selten for their insightful teachings.

1. The Conceptual Framework

Our generalized representation of a two-person cooperative game mainly redefines the payoff function in an innovative way conceptually. The conventional payoff function in economics and in game theory is typically interpreted as a "representation" of the player's "all-things-considered" preferences over outcomes: "I get a payoff of 5 from the outcome $x$ and a payoff of 1 from the outcome $y$" is equivalent to, "all things considered, I rank $x$ higher than $y$, or strictly prefer $x$ to $y$". Normally, no assumptions are made regarding the motives underlying a player's ranking or preferences over the outcomes (see Sen 1977); "what is being assumed is that the agents who have multiple cares and concerns have resolved any conflicts into an 'all-things-considered preferences'" (Baigent 1995, p. 92). In this paper, we shall attempt to tease out different motives/values out of the payoff function.[i]

Preferences play a dual role in economics and in game theory: on the one hand, they are used to explain observed behaviors and to predict future behaviors, and on the other hand, they are regarded as indicators of a player's welfare. Accordingly, being a representation of a player's preferences, a player's payoff function has similar roles. von Neumann and Morgenstern (1944, 2004 edition, p.8-9) interpreted the payoff of a player as his/her "utility or degree of satisfaction" and proposed, as the "notion of rationality", the player's utility will be maximized. The interpretation of the payoff of a player as the degree of the player's preference satisfaction (or desire fulfilment) is in line with the conventional interpretation of utility in modern microeconomic theory. However, as noted earlier, a player's preferences are taken as the player's "all-things-considered" preferences, and, as such, the degree of preference satisfaction may not capture the player's utility-based conception of well-being/welfare adequately and may lead to misleading welfare analysis. The philosophical literature on welfare economics points to several problems of the conception of welfare based on the preference satisfaction of a player's preferences (see, among others, Elster (1983), Goodwin (1995), Hahn (1982), Pattanaik and Xu (2024), Sen (1985)): changing preferences, preferences based on wrong and/or inadequate factual information, multiple and conflicting preferences, antisocial and malicious preferences, adaptive preferences, etc., to name a few.

Faced with such problems, Harsanyi (1977a, 1977b) has suggested of distinguishing between the actual preferences of a player and the player's true preferences as they would be under ideal conditions; indeed, Harsanyi (1977a, p.29-30) notes that "It is well-known that a person's preferences may be distorted by factual errors, ignorance, careless thinking, rash judgment, or strong emotions hindering rational choice, etc.", and suggests that "Therefore, we may distinguish between a person's *explicit* preferences, i.e., his preferences as they actually *are* … and his `true' preferences, i.e., his preferences as they *would* be under `ideal conditions'."

In this paper, we shall follow Harsanyi's suggestion to examine and redefine a player's payoff. For a player $i$, let $i$'s payoff (representing the player's actual preferences), $p_i$, consist of two parts: Part one is the "utility" payoff $U_i$ (representing the player's "true" preferences; to be called "the rational value function" following von Neumann and Morgenstern, 2004 edition), while Part two is the "distortion" payoff $D_i$ (reflecting the part of the player's non-true preferences; to be called "the distortion value function"). The distortion value function is also a mapping of all *n*-tuples of pure or mixed strategies into the real numbers where $n$ is the number of players in the game, with $n = 2$ in this paper, and represents what the player will bring into the game in a sense of "irrationality", rather than any "true-preference" satisfaction.[ii] For example, in the present China, there are frequent tragic events that some individuals go to streets to hurt or even kill people randomly whom he/she did not know at all (particularly by driving a car or a truck). Such behavior is totally irrational from any perspective of traditional "utility". Rather, such act of destruction seems to be a leak of "anger" for destruction *per se*. We shall assume the "value" of such irrationality can also be expressed by real numbers, in line with the utility measurement, for simplicity.

The idea of "irrationality" mentioned above is not about intransitive preferences of a player; rather, it refers to some particular behaviors or concerns or values that are deemed to distort the player's "true" preferences and/or sometimes to be destructive to both the player himself/herself and others—behaviors that harm others without benefiting oneself.

Our approach may thus be viewed as making a suggestion for a research paradigm shift: we should use the payoff function $p_i$ to understand and predict the behaviors of a player $i$, but when it comes to analyze $i$'s welfare, we should use $i$'s rational value function/utility payoff $U_i$. Thus, the alleged "burden" of the dual role of a player $i$'s "all-things-considered" preferences can be divided and shared between $p_i$ and $U_i$: to explain the behavior, we should use $i$'s payoff function $p_i$, and, to conduct welfare analysis, we should use $i$'s rational value function $U_i$.

We now formulate a player's payoff formally. For simplicity, we take that the payoff function $p_i$ of a player $i$ is given by $p_i(s_1, s_2) = \delta_i U_i(s_1, s_2) + (1 - \delta_i) D_i(s_1, s_2)$, with $s_i$ as a strategy (pure or mixed) of player $i$ and where $\delta_i \in [0,1]$ with $\delta_i$ being interpreted as player $i$'s "degree of rationality". When $\delta_i = 1$ for all $i$, we are back to the traditional game with all the rational analysis of behaviors and conventional welfare analysis. When $\delta_i = 0$ for all $i$, this game will go to the other extreme of "full irrationality" such as everybody going out to hurt unknown people randomly. In between, when $0 < \delta_i < 1$ for some $i$, we have a game of "limited rationality", more or less. Therefore, one may call this index $\delta$ as an index of rationality, in a general sense.

Following Nash (1953, p.130-131), we outline briefly the formal negotiation model (the Nash demand game) where the cooperative game is reduced to a non-cooperative game. Recall that in the Nash demand game, two players have the opportunity to divide a certain amount of money between them.

*Stage one*: Each player ($i$) chooses a strategy $t_i$ (threat) which he/she will be forced to use if the two cannot come to an agreement, that is, if their demands are incompatible.

*Stage two*: The players inform each other of their threats.

*Stage three*: Each player decides upon his/her demand $d_i(\delta_i)$, which is a point on his payoff scale $p_i(s_1, s_2) = \delta_i U_i(s_1, s_2) + (1 - \delta_i) D_i(s_1, s_2)$, according to his/her rationality level $\delta_i$. Note that it is not the utility scale in Nash (1953).

*Stage four*: The payoffs are now determined. If there is a point $(p_1, p_2)$ in the payoff space such that $p_1 \geq d_1(\delta_1)$ and $p_2 \geq d_2(\delta_2)$, then the payoff to each player $i$ is $d_i(\delta_i)$. That is, if the demands can be simultaneously satisfied, then each player gets what he/she demanded. Otherwise, the payoff to player $i$ is $p_i(t_1, t_2)$; i.e., the threats must be executed.

**Theorem 1:** The solution of this two-person cooperative game with δ-rationality is $(d_1(\delta_1), d_2(\delta_2))$.

Proof: Analogous to Nash (1953) p. 131-136, with $d_1(\delta_1)$ and $d_2(\delta_2)$ instead of $d_1$ and $d_2$.

**Remark 1.** Every generalized two-person cooperative game with δ-rationality has a Nash solution point for each $\delta_i \in [0,1]$. Thus, there are a possible continuum of Nash solution points for various rationality levels.

**Remark 2.** If $\delta_i = \bar{\delta}$ (a constant) for all *i*, the generalized game has Nash solution points all of which as a function of the constant rationality index $\bar{\delta}$. If all players have the same rationality level $\bar{\delta}$, we shall call such a generalized game as a two-person cooperative game of constant-rationality. The extreme cases are $\bar{\delta}=1$, back to the traditional game type with the standard Nash solution, and $\bar{\delta}=0$, the fully irrational game.

**Remark 3.** The distribution characteristics of $\delta_i$, when $\delta_i$ is viewed as a random variable on [0, 1], will set distribution characteristics of solution points. We shall call such a generalized game as a two-person cooperative game of variable-rationality game. For example, if $\delta_i \in [0,1]$ follows a normal distribution for all *i*, we may call such a generalized game as a two-person cooperative game of normal distribution rationality.

**Remark 4.** The solution $(d_1(\delta_1), d_2(\delta_2))$ can be viewed as a behavioral outcome resulting from the players' actual preferences (all-things-considered preferences) as represented by their payoffs. To analyze the welfare properties of the solution, one would need to use the players' "true preferences" as represented by their rational value functions.

## 2. A Numerical Example

For illustrative purpose, we shall give a simple numerical example as follows.

"Suppose two players are to divide between them a potential profit of $100. If the players come to an agreement, they divide the money based on their agreement; if they fail to agree, neither of them receives anything." Maschler et al. (2013), p.623, Example 15.1.

Let $s_i$ be the amount of the money that player $i$ receives. Suppose that Player 1's and 2's utility functions are $U_1(s_1, s_2) = s_1$, $U_2(s_1, s_2) = s_2$. It is well known that the Nash solution is $(d_1, d_2)$=(50, 50).

Suppose now that Player 1's and 2's distortion value functions are $D_1(s_1,s_2) = s_1 - s_2$, $D_2(s_1,s_2) = s_2 - s_1$, $0 \leq s_1, s_2 \leq 100$, with an intuitive interpretation that the players would be jealous when the money he/she received is less than the money the other player received. That is, everybody wants to have more portion and somewhat uncomfortable with the cut by the other player, which is not unimaginable in real business cases as we are informed by practitioners.

Suppose that Player 1 and 2's rationality levels are $\delta_1, \delta_2 \in [0,1]$ respectively. Then, the payoff functions of Players 1 and 2 become
$p_1(s_1,s_2) = \delta_1 s_1 + (1 - \delta_1)(s_1 - s_2)$ and $p_2(s_1,s_2) = \delta_2 s_2 + (1 - \delta_2)(s_2 - s_1)$,
respectively, and the payoff point for the disagreement is $(0,0)$.

The division $(s_1^*, s_2^*)$ corresponding to the solution $(d_1(\delta_1), d_2(\delta_2))$ solves the problem:

$max_{(s_1,s_2)}(\delta_1 s_1 + (1 - \delta_1)(s_1 - s_2))(\delta_2 s_2 + (1 - \delta_2)(s_2 - s_1))$ s.t. $0 \leq s_1, s_2 \leq 100$ and $s_1 + s_2 \leq 100$.

Then, it can be checked that,

if $0 < \delta_1, \delta_2 \leq 1$, then $(p_1(s_1^*, s_2^*), p_2(s_1^*, s_2^*)) = (\frac{50(\delta_1+\delta_2-\delta_1\delta_2)}{2-\delta_2}, \frac{50(\delta_1+\delta_2-\delta_1\delta_2)}{2-\delta_1})$, which

corresponds to $(s_1^*, s_2^*) = (\frac{100(1-\delta_1)}{2-\delta_1} + \frac{50(\delta_1+\delta_2-\delta_1\delta_2)}{(2-\delta_1)(2-\delta_2)}, \frac{100(1-\delta_2)}{2-\delta_2} + \frac{50(\delta_1+\delta_2-\delta_1\delta_2)}{(2-\delta_1)(2-\delta_2)})$ (see

Diagram 1), and,
if $\delta_1 = 0$ or $\delta_2 = 0$, then $(p_1(s_1^*, s_2^*), p_2(s_1^*, s_2^*)) = (0,0)$ so that the disagreement point prevails.

It may be noted that, when $0 < \delta_1, \delta_2 \leq 1$, we have $s_1^* > s_2^*$ if and only if $\delta_1 < \delta_2$. Thus, the solution suggests non-equal division of the money in general and that the "more rational" player would get less money than the "less rational" player.

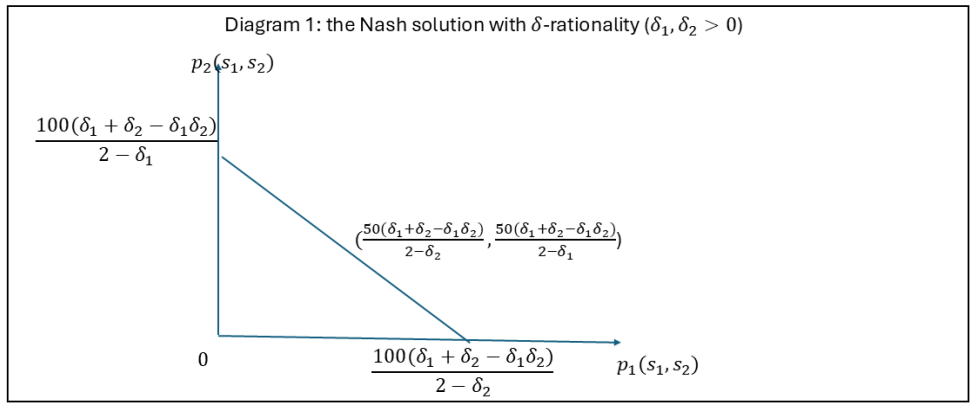

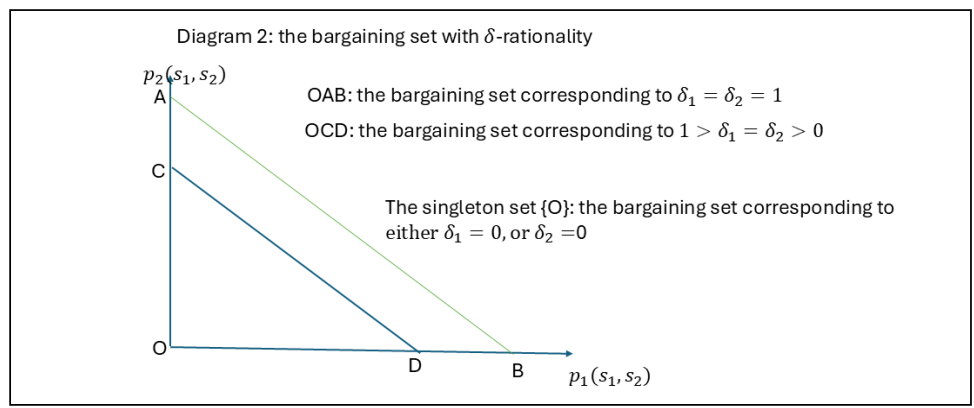

It may also be noted that, (1) when $\delta_1 = \delta_2 = 1$, the "bargaining set" (the set of options that are feasible and Pareto dominate the disagreement point) is the largest, (2) when both $\delta_1$ and $\delta_2$ decrease, the bargaining set shrinks (so that the "profitable opportunities" feasible and available to both players become less), and (3) in particular, when either player becomes totally "irrational" (either $\delta_1 = 0$ or $\delta_2 = 0$), the bargaining set is reduced to the singleton set containing the disagreement point as the only non-dominant point (suggesting the opportunity for cooperation is reduced to nil). Cooperation creates value, value needs cooperation.

Needless to say, more empirical and experimental investigation is certainly needed in the future, as well as further theoretical exploration.

### 3. A Real-Life Business Case

We translate a brief summary by the Founder & CEO of an AI application start-up, word by word, as follows. This Founder & CEO and the lawyer of this case can be contacted through us, upon request, for any field study purpose.

In February 2023, there were two parts of initiators in an AI application start-up: A (Founder CEO, technology + product + management) 55%, B (multiple individuals, capital partners, investment RMB 3 million + market) 45%. This is the initial partnership which is also consistent with the industry convention that one party supplies technology and product capability and the other party supplies capital.

During the development process of this start-up, investor C was introduced to join. After the participation of C as a shareholder through investment, the shareholding structure of this start-up became: A (Founder CEO, technology + product + management) 44%, B (multiple individuals, financial partners, investment + market) 36%, C (investment shareholder, investment of RMB 3 million), 20%.

In December 2023, for further development, the start-up needed to attract more investment, and then C intended to invest more but had one requirement: All shareholders put in more investment according to the shareholding ratio.

Then, shareholder A as the founder and technology product input partner disagree with this requirement.

Shareholder B, composed by multiple individuals, exhibited different opinions: B1 firmly opposed this requirement and viewed this as violation of initial agreement; B2 viewed this requirement as deeply bonding the founder thus more beneficial for the growth of the start-up, and since he/she was a minor shareholder, his/her would contribute less money due to the requirement of the founder contributing money investment, benefits might become even bigger; B3 took a standby position to follow the majority of stakeholders.

By that time, the originally-united shareholders split.

By that time, for the external environment, AI was like a sun shining in the sky, this start-up obtained the AI business license by the Cyberspace Administration of China (only about 100+ companies obtained this license in Mainland China by that time – now several hundreds), with Shenzhen Huawei and Tencent in the same batch. This start-up's product had won the Top 50 AI Application Award in the industry evaluation, and led the technology breakthrough and obtained the invention patent (the product technology emerging in this industry after 1 year proved the leading role of this start-up's technology and product and its accurate market prediction). Multiple investment organizations were following the development of this start-up, and the due diligence work was undergoing. When the information of disagreement among the shareholders was known by the external outside investment parties with investment intentions, those preparing investors viewed the shareholding structure of this start-up as unreasonable and also due to the existence of internal conflicts, all those preparing investors chose to pause.

By that time, the Founder CEO firmly believed that this start-up had a future, therefore raised funds himself to maintain the operation of this start-up, for a win-win strategy to keep the company alive and the product should continue to operate. He also proposed to give up most shares to let the internal shareholders have the priority of investments, to maintain the company alive.

Then C asked to be the major shareholder but did not clarify the proportion as the major shareholder. C played by delay and passing the buck etc. to let the maintenance pressure of this company further on the Founder CEO, to force the founder A give in.

By that time, shareholders B believed that if the founder chose to quit, the start-up would certainly die, therefore they chose to persuade C clarifying his position. C had all the initiatives then.

Under optimal conditions, C proposed to have 80% share of the stat-up by RMB 1 yuan, after the start-up spent the last loan through the founder A.

The founder A immediately called for the shareholder meeting and asked all shareholders to vote for the solutions for the start-up, all the process being audio-recorded by a lawyer as the legal process required. B2 chose not to attend the shareholder meeting, B1 and B3 chose to vote with A for bankruptcy liquidation, then there were more than 2/3 shareholder votes. C voted against this solution but could not change the decision of the shareholder meeting, because his 20% was not enough to cast a veto.

By that time, C proposed to be the good guy to take over the company, not to let the company die, and to renegotiate a new deal, since the start-up would be liquidated and die. A proposed to quit from all interests and be willing to fully cooperate in the work handover and then no more support. A hoped to main his social credibility, to take all the historic debt by this company, all the company assets to be transferred to all the C and B shareholders by RMB 1 yuan, and all the shares of C and B to be transferred to A, that is, a solution of transferring all the assets but maintain the shell company. In the meantime, A proposed that this solution still needed to go through the shareholder meeting, to formulate a formal document, and set the deadline of the decision time, that is, if there was no unanimous agreement within an agreed-upon time framework then the already formed bankruptcy liquidation decision of the shareholder meeting must be executed.

After this discussion, C proposed that all the assets be transferred to a company by C, but C's shares be kept, that is, C wanted all the assets and all the shell. More specifically, C wanted all the valuable assets of the company such as the patents, trademark and source code etc. and to keep the 20% shares but not all the shares of the company because the debt of the company was bonded with the company. C would not take all the company shares and the corporation legal person status thus C could put all the company debt burden to A. If C took all the company shares, A could exit freely. C chose to keep his 20% shares to contain and pester A as an effective restriction on A for a new start-up.

Perhaps C thought that A wanted to keep the company shell because there were more valuable things, but A actually only wanted to maintain his social credibility. Or perhaps C thought that simply not to let A be satisfactory, not to execute A's strategy, that is, "whatever the enemy supports, I will oppose".

By October 2024, A thought that there was no more necessity for any communications, and directly submitted the bankruptcy liquidation file to the court, to close the AI service of this start-up, the start-up fell completely down.

A regarded this failure as the biggest one in his entrepreneurship career, and in this case, all stakeholders lost. A thought that C was the biggest loser financially, and A himself was the biggest loser overall due to the enormous overall loss and opportunity costs.

Months after this litigation, A had discussed this case with one of the most senior investors in China and received the pacifying but sad response that "70% of the failed start-ups were due to the similar reasons, more or less", in this senior investor's life-time experiences with the start-ups he had contacts.

---

[i] There is a huge literature in experimental and behavioral economics that has attempted to investigate various concerns/motives lumped in a player's payoff function; see, for example, Fehr and Schmidt (2006) for an account of such concerns/motives.

[ii] Max Weber (1922; 1978, p. 6) remarked that "The more we ourselves are susceptible to such emotional reactions as anxiety, anger, ambition, envy, jealousy, love, enthusiasm, pride, vengefulness, loyalty, devotion, and appetites of all sorts, and to the "irrational" conduct which grows out of them, the more readily can we empathize with them. … By comparison with this it is possible to understand the ways in which actual action is influenced by irrational factors of all sorts, such as affects and errors, in that they account for the deviation from the line of conduct which would be expected on the hypothesis that the action were purely rational."